%% file: paper.tex
\documentstyle{mn}
\input{psfig}
\input{macros_vdbosch}
\begin{document}


\title[The Mass Function and Average Mass-Loss Rate of Dark Matter Subhaloes]
      {The Mass Function and Average Mass-Loss Rate of Dark Matter Subhaloes}
\author[van den Bosch, Tormen \& Giocoli]
       {Frank C. van den Bosch$^{1}$, Giuseppe Tormen$^{2}$ and 
        Carlo Giocoli$^{2}$
       \thanks{E-mail: vdbosch@phys.ethz.ch}\\
        $^1$Department of Physics, Swiss Federal Institute of
            Technology, ETH H\"onggerberg, CH-8093, Zurich,
            Switzerland\\
        $^2$Dipartimento di Astronomia, Universit\`a di Padova,
         Vicolo dell`Osservatorio 2, I-35122 Padova, Italy} 


\date{}

\pagerange{\pageref{firstpage}--\pageref{lastpage}}
\pubyear{2000}

\maketitle

\label{firstpage}


\begin{abstract}
  We  present a  simple,  semi-analytical model  to  compute the  mass
  functions of dark matter subhaloes. The masses of subhaloes at their
  time of accretion  are obtained from a standard  merger tree. During
  the subsequent evolution, the  subhaloes experience mass loss due to
  the  combined effect  of  dynamical friction,  tidal stripping,  and
  tidal  heating.    Rather  than  integrating   these  effects  along
  individual subhalo  orbits, we consider the average  mass loss rate,
  where the average is taken over all possible orbital configurations.
  Under  the  ansatz  that  the  average  distribution  of  orbits  is
  independent of parent halo mass, this allows us to write the average
  mass loss  rate as a simple  function that depends  only on redshift
  and  on the instantaneous  mass ratio  of subhalo  and parent  halo. 
  After calibrating  this model by matching the  subhalo mass function
  (SHMF)   of   cluster-sized  dark   matter   haloes  obtained   from
  high-resolution, numerical simulations, we investigate the predicted
  mass and redshift dependence of the SHMF.  We find that, contrary to
  previous  claims,  the  subhalo  mass  function is  not  universal.  
  Instead, both the slope and the normalization depend on the ratio of
  the parent  halo mass, $M$,  and the characteristic  non-linear mass
  $M^{*}$. This simply reflects a halo formation time dependence; more
  massive parent haloes  form later, thus allowing less  time for mass
  loss  to  operate.  We  predict  that  galaxy-sized  haloes, with  a
  present-day mass of $M \simeq  10^{12} h^{-1} \Msun$ have an average
  mass fraction of dark matter  subhaloes that is a factor three lower
  than for massive clusters with  $M \simeq 10^{15} h^{-1} \Msun$.  We
  also analyze  the halo-to-halo scatter  in SHMFs, and show  that the
  subhalo mass fraction of  individual haloes depends most strongly on
  their accretion history in the last $\sim 1$ Gyr. Finally we provide
  a simple fitting  function for the average SHMF of  a parent halo of
  any mass at any redshift  and for any cosmology, and briefly discuss
  several implications of our findings.
\end{abstract}




\section{Introduction}
\label{sec:intro}

During  the hierarchical  assembly of  dark matter  haloes,  the inner
regions  of early virialized  objects often  survive accretion  onto a
larger system,  thus giving rise  to a population of  subhaloes.  This
substructure  evolves as it  is subjected  to the  forces that  try to
dissolve   it:  dynamical  friction,   tidal  forces,   and  impulsive
collisions.   Depending  on  their  orbits  and  their  masses,  these
subhaloes  therefore either  merge, are  disrupted or  survive  to the
present day.

To fully describe, in a statistical sense, the non-linear distribution
of mass  in the  Universe, it is  essential that halo  substructure is
taken into account.  After all,  galaxies are thought to reside at the
centers of dark  matter haloes, which includes dark  matter subhaloes. 
When  building a  coherent picture  of galaxy  formation or  of galaxy
clustering,  it  is  therefore   of  paramount  importance  that  halo
substructure is taken into account. In particular, we need an accurate
description of  the conditional subhalo  mass function, $n(m  \vert M)
{\rm  d}m$, which gives  the number  of subhaloes  with masses  in the
range $m  \pm {\rm d}m/2$ that  reside in a  parent halo of mass  $M$. 
Combined with the  (parent) halo mass function, $n(M)  {\rm d}M$, this
then provides a complete,  statistical description of the abundance of
dark  matter haloes down  to the  level of  subhaloes. In  addition, a
comparison of $n(m \vert M)  {\rm d}m$ with the conditional luminosity
function, $\Phi(L \vert M) {\rm d}L$  (Yang, Mo \& van den Bosch 2003;
van den  Bosch, Yang  \& Mo 2003)  will yield important  insights into
galaxy formation and allow for a detailed study of galaxy bias.

Only  since  a couple  of  years  numerical  simulations of  structure
formation have  reached the mass and  force resolution to  allow for a
detailed study of dark matter substructure (e.g., Tormen 1997; Tormen,
Diaferio \& Syer  1998; Moore \etal 1998, 1999;  Klypin \etal 1999a,b;
Ghigna  \etal 1998,  2000; Stoehr  \etal  2002; De  Lucia \etal  2004;
Diemand, Moore  \& Stadel  2004; Gill \etal  2004a,b; Gao  \etal 2004;
Reed \etal  2004; Kravtsov  \etal 2004).  Most  of these  studies have
found  that in  terms of  their substructure  properties,  dark matter
haloes are  homologous; the internal structure of  a galaxy-sized halo
looks just  like a rescaled version  of that of a  rich cluster.  This
would imply  that the subhalo  mass function is independent  of parent
halo mass.  However, most of  these results are based on small numbers
of individual haloes, while halo-to-halo variations are expected to be
fairly large.  Combined with uncertainties due to numerical resolution
and the identification  of dark matter subhaloes, this  means that the
statistical  significance  of these  results  is  still unclear.   For
example, Gao \etal (2004), analyzing a relatively large sample of dark
matter haloes extracted from a large, high-resolution simulation, find
that the  normalization of the SHMF  depends on parent  halo mass.  In
particular, they claim that more  massive haloes contain a larger mass
fraction in  subhaloes. Similar results have been  obtained by Diemand
\etal (2004; their Fig.~7) and Kang \etal (2004; their Fig.~2). Such a
parent halo mass dependence might  be expected from the fact that more
massive haloes  form later  (e.g., van den  Bosch 2002),  thus leaving
less  time for mass  loss to  operate. 

Recently, there have also been  a number of analytical studies of dark
matter subhaloes based on the extended Press-Schechter (EPS) formalism
(Bond \etal 1991;  Bower 1991; Lacey \& Cole  1993).  Although the EPS
formalism  only yields information  regarding parent  haloes, it  is a
logical next step to simply associate the progenitor haloes of a given
parent  halo (whose  properties can  be computed  using EPS)  with its
present day subhaloes (Fujita \etal 2002; Sheth 2003).  This, however,
ignores the fact that subhaloes experience significant amounts of mass
loss.  A more realistic approach, therefore, needs to combine this EPS
based formalism with an analytical description of the (mass) evolution
of dark matter subhaloes.

Oguri \& Lee  (2004), following up on a previous  study by Lee (2004),
presented a semi-analytical model to compute the SHMF from EPS, taking
detailed  account of  dynamical  friction and  tidal stripping.   They
predict that  the SHMF is virtually  independent of parent  halo mass. 
However, an  obvious downside of their  approach is that  they use the
present  day mass  of the  parent halo  when computing  the  impact of
dynamical friction.   In reality, the parent halo  mass evolves, which
should have  been taken  into account (see  e.g., Taffoni  \etal 2003;
Zhao  2004).   Therefore, it  seems  likely  that  Oguri \&  Lee  have
underestimated the  impact of dynamical friction, and,  since the mass
accretion history  depends on mass,  may not have  correctly predicted
the mass-dependence of the  SHMF.  Zentner \& Bullock (2003; hereafter
ZB03) and  Taylor \& Babul (2004;  hereafter TB04) improve  on this by
integrating orbits in the changing potential of the parent halo (whose
mass accretion  history is computed using detailed  merger trees).  By
including detailed  analytical descriptions of  dynamical friction and
tidal heating and stripping, these authors provide detailed, realistic
models for the evolution  of dark matter substructure.  Unfortunately,
TB04  refrain from  a discussion  of predictions  regarding  the SHMF,
while  ZB03 only  investigate  the cosmology-dependence,  but not  the
parent halo mass dependence.

In this paper  we follow a similar approach, except  that we treat the
actual  mass  loss of  subhaloes  in a  very  simple  manner. We  only
consider  the {\it  average}  mass  loss rate,  where  the average  is
considered to  be taken over  all orbital configurations.   This means
that we do  not have to integrate individual orbits,  and allows us to
write the mass loss rate as a function of the mass ratio of subhalo to
parent halo  only.  Rather  than attempting to  obtain an  estimate of
this average mass  loss rate from first principles,  we simply adopt a
functional  form, and  adjust the  free parameters  to match  the SHMF
obtained from  numerical simulations.   As in ZB03  and TB04,  we take
detailed account of  the fact that while the  subhalo looses mass, the
parent  halo  gains  mass  due  to  its  hierarchical  growth.   After
calibrating  the model  against numerical  simulations, we  use  it to
investigate the parent halo mass  and redshift dependence of the SHMF,
as well  as the halo-to-halo scatter.   We show that  our simple model
predicts that (i)  more massive haloes have a  larger mass fraction of
substructure,  (ii)  the  halo-to-halo  scatter is  large,  (iii)  the
abundance of subhaloes {\it per  unit parent halo mass} is independent
of parent mass, and (iv) the subhalo mass fraction is larger at higher
redshifts.   These  findings  are  in  excellent  agreement  with  the
numerical simulations of Gao \etal  (2004).  The main advantage of our
model over either numerical simulations or the more detailed models of
ZB03 and  TB04 is  its shear simplicity  and computational  speed that
allows  a detailed  investigation of  the  dependence of  the SHMF  on
cosmology, parent halo mass, and redshift.  In addition, it provides a
simple  description of  the  average  mass loss  rate  of dark  matter
subhaloes, which may be useful, for example, to describe the evolution
of the mass-to-light ratio of satellite galaxies.

This  paper is  organized as  follows. In  Section~\ref{sec:numres} we
give  a   brief  overview  of   the  SHMFs  obtained   from  numerical
simulations.    Section~\ref{sec:theory}  describes  our   method  for
computing SHMFs based  on a combination of EPS and  a simple model for
the average mass  loss rate of subhaloes.  Sections~\ref{sec:massdep},
\ref{sec:scatter}, and~\ref{sec:redshift} discuss the mass-dependence,
the halo-to-halo  variance, and the  redshift dependence of  the SHMF,
respectively.  In Section~\ref{sec:fit} we provide a simple analytical
fitting function  for the  average SHMF  of a halo  of given  mass and
redshift.  We summarize our results in Section~\ref{sec:concl}.

Throughout we use $m$ and $M$  to denote the masses of the subhalo and
the parent halo,  respectively.  Here the parent halo  mass is defined
as the total mass (including that of all subhaloes) within a sphere of
density 200 times the critical  density at redshift zero.  For brevity
we use  $\psi$ to indicate  the mass ratio  $m/M$, and we  consider it
understood  that $m$,  $M$, and  $\psi$  all depend  on time,  without
having to write this  time-dependence explicitly.  A subscript zero is
used  to indicate  the present  day value  (i.e., at  redshift  zero). 
Unless  specifically stated  otherwise, we  adopt a  flat $\Lambda$CDM
`concordance'  cosmology with  $\Omega_m=0.3$, $\Omega_{\Lambda}=0.7$,
$h=H_0/(100  \kmsmpc) =  0.7$  and with  initial density  fluctuations
described  by  a  scale-invariant  power spectrum  with  normalization
$\sigma_8=0.9$.

\section{Subhalo Mass Functions from Numerical Simulations}
\label{sec:numres}

\begin{figure}
\centerline{\psfig{figure=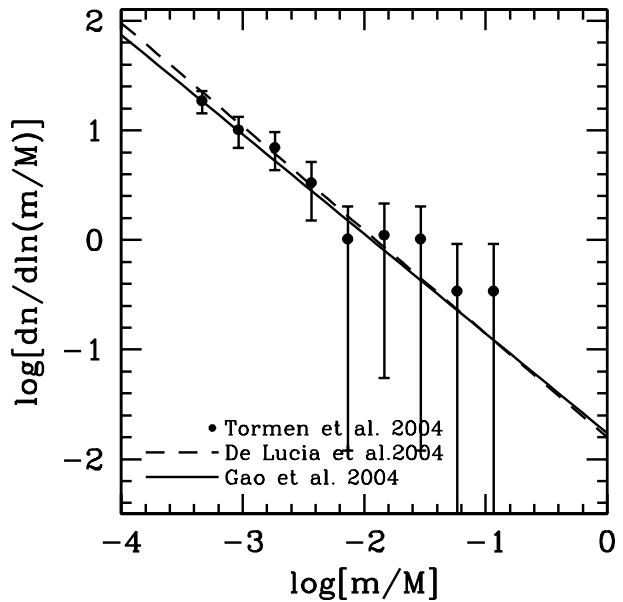,width=\hssize}}
\caption{Comparison of  SHMFs of parent haloes in the mass range
  $10^{14}  h^{-1}  \Msun \lta  M_0  \lta  10^{15}  h^{-1} \Msun$,  as
  obtained  by  different  authors  using  different  high  resolution
  numerical simulations.}
\label{fig:simres}
\end{figure}

As  discussed in  the introduction,  numerous studies  have determined
SHMFs     from    high-resolution    numerical     simulations.     In
Fig.~\ref{fig:simres} we compare the subhalo mass functions from three
independent studies  (all based  on the same  $\Lambda$CDM concordance
cosmology).  The  solid dots with errorbars  (Poissonian) indicate the
average  SHMF  obtained using  the  simulations  described in  Tormen,
Moscardini \& Yoshida (2004) from 17 clusters with masses in the range
$3 \times 10^{14} h^{-1} \Msun \leq M_0 \leq 1.7 \times 10^{15} h^{-1}
\Msun$.   These high-resolution  simulations were  obtained  using the
technique  of re-simulating, at  much higher  resolution, a  region of
interest selected  from a large  cosmological volume.  De  Lucia \etal
(2004) studied a  similar set of 11 high  resolution re-simulations of
galaxy  clusters with  masses in  the range  $7 \times  10^{13} h^{-1}
\Msun \leq M_0 \leq 1.8 \times 10^{15} h^{-1} \Msun$.  The dashed line
in Fig.~\ref{fig:simres} corresponds to ${\rm d}n/{\rm d}{\rm ln}(m/M)
= 0.016 (m/M)^{-0.94}$, which is the power-law relation that best fits
the average  SHMF of  this set (obtained  by fitting their  results by
eye).\footnote{The normalization of the  mass functions shown in panel
  (f) of Fig.~1 of De Lucia  \etal (2004) is incorrect and needs to be
  translated  in  the  $y$-direction  by $+0.6$,  (De  Lucia,  private
  communication)}.   Finally, the solid  line indicates  the power-law
SHMF, ${\rm d}n/{\rm d}{\rm ln}(m/M) = 0.017 (m/M)^{-0.91}$, that best
fits the results  of Gao \etal (2004), obtained  by fitting-by-eye the
average SHMF  of their 15 haloes  with $3 \times  10^{14} h^{-1} \Msun
\leq M_0 \leq 10^{15} h^{-1} \Msun$.

All three SHMFs  are in good agreement with each  other, both in terms
of the slope at small $m/M$ and in terms of the normalization.  In the
range $-3.5 \leq {\rm log}(m/M) \leq -2.5$, where the results are most
accurate,  all three SHMFs  agree with  each other  at better  than 20
percent.   Given the different  force and/or  mass resolutions  of the
various  simulations, and  the different  techniques used  to identify
subhaloes, this  level of  agreement is in  fact better than  what one
might naively  expect.  Especially  since relatively small  samples of
haloes have  been used, which,  if halo-to-halo scatter is  large, may
cause     significant    scatter     in     these    averages.      In
Section~\ref{sec:evolve} below  we will  use these SHMFs  to calibrate
our model for the subhalo mass loss rate.

\section{Subhalo Mass Functions from Merger Trees}
\label{sec:theory}

The aim  of this paper is to  develop an algorithm that  allows a fast
and reliable  computation of subhalo  mass functions. As  discussed in
Section~\ref{sec:intro}  two ingredients  are essential:  a  method to
compute  progenitor  haloes,  and  a  proper  treatment  of  the  mass
evolution of subhaloes. For the former, we use a standard merger tree,
which  we  construct  using   the  $N$-branch  method  with  accretion
(Somerville  \&  Kolatt  1999;  hereafter  SK99). We  adopt  the  same
time-stepping  as  in SK99,  and  introduce  a  lower-mass cut-off  of
$m_{\rm  lim}  =  10^{-4}  M_0$,  which reflects  the  effective  mass
resolution of  our merger trees.   This cut-off is required  since the
number  of  progenitor haloes  diverges  as the  mass  goes  to zero.  
Following  SK99, any  mass contained  in haloes  below  the resolution
limit  is accounted for  by referring  to it  as `accreted'  mass (for
which the prior mass accretion history is not followed back in time).
\begin{figure}
\centerline{\psfig{figure=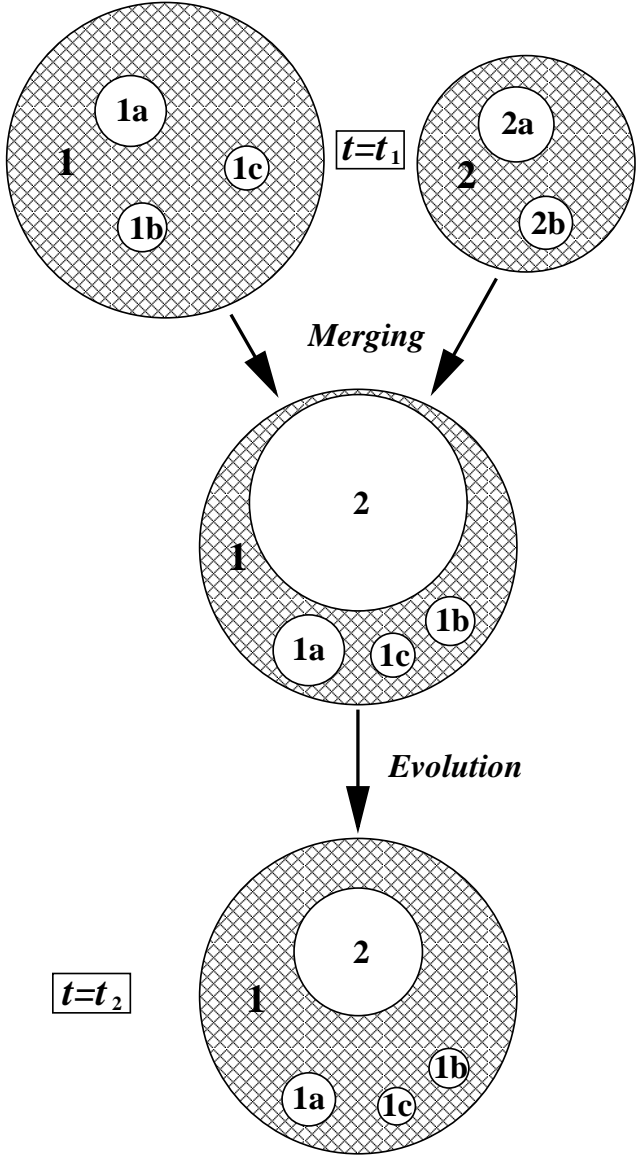,width=\hssize}}
\caption{Illustration of the processes of merging and evolution. Each
  time-step, such as the step  between $t=t_1$ and $t=t_2$ shown here,
  the  parent halo  mass grows  by merging  (assumed  instantaneous at
  $t=t_1$), while the subhalo masses evolve due to mass loss. See text
  for a detailed description.}
\label{fig:drawing}
\end{figure}

A  proper treatment of  the mass  evolution of  the subhaloes  is more
complicated.   A subhalo  moving on  a fixed  orbit in  a  static halo
experiences  mass loss  due to  tidal stripping  and heating.   If the
orbit were to  remain fixed, and in the absence  of tidal heating, the
mass loss rate would rapidly decline with time, as all mass beyond the
tidal radius  would be stripped after  at most a few  orbital periods. 
In reality,  however, tidal heating  continues to `push'  stars beyond
the tidal radius, where they  can be stripped, while the orbit evolves
due to  dynamical friction  which causes the  tidal radius to  shrink. 
Both  effects  significantly prolong  the  duration  and increase  the
amount  of  mass loss,  which  may  eventually  lead to  the  complete
disruption of the subhalo.   For detailed numerical simulations of the
mass  loss of  dark  matter  subhaloes see  Hayashi  \etal (2003)  and
Kazantzidis \etal (2004).

To  properly account  for the  above mentioned  effects,  which depend
strongly on the orbital  eccentricity (e.g., Colpi, Mayer \& Governato
1999; Gnedin, Hernquist \& Ostriker  1999; Taylor \& Babul 2001, 2004;
Taffoni  \etal  2003),  requires   a  detailed  integration  over  all
individual subhalo  orbits. This is  complicated by the fact  that the
mass of the parent halo evolves with time.  If the mass growth rate is
sufficiently  slow,  the  evolution  may be  considered  an  adiabatic
process,  thus  allowing the  orbits  of  subhaloes  to be  integrated
analytically despite the non-static nature of the background potential
(this  principle is exploited  in the  models of  ZB03 and  TB04).  In
reality, however, haloes  grow hierarchically through (major) mergers,
making the actual orbital evolution highly non-linear.

In order to sidestep these difficulties, we consider the {\it average}
mass loss  rate of dark matter  subhaloes, where the average  is to be
taken over  the entire  distribution of orbital  configurations.  This
removes  the  requirement  to  actually integrate  individual  orbits,
resulting in  a subhalo mass loss  rate that depends only  on the mass
ratio $\psi  = m/M$\footnote{Throughout this paper we  ignore the fact
  that the average  halo concentration depends weakly on  halo mass.}. 
This  is  true  as  long   as  the  average  distribution  of  orbital
eccentricities of subhaloes does not  depend on parent halo mass.  The
eccentricity  distribution,  $P(\epsilon)$,  depends  on  the  orbital
anisotropy and on the density distribution of the parent halo (van den
Bosch \etal 1999).  Although more massive haloes are less concentrated
on average (Navarro, Frenk \& White  1997), which could give rise to a
mass  dependence of  $P(\epsilon)$, van  den Bosch  \etal  (1999) have
shown  that $P(\epsilon)$ depends  much more  strongly on  the orbital
anisotropy than on the actual density distribution of the parent halo.
Since there is  no obvious reason why the anisotropy  of the orbits of
subhaloes should depend on halo mass, our assumption that the average,
instantaneous  mass loss rate  of substructure  depends only  on $m/M$
should be sufficiently accurate.

\subsection{The average mass loss rate}
\label{sec:massloss}

During the evolution of the system the parent mass, $M$, will increase
due  to merging  and accretion,  whereas the  subhalo mass,  $m$, will
decrease due to  the effects discussed above.  We  postulate that in a
steady-state halo, for  which $\dot{M} \equiv {\rm d}M/{\rm  d}t = 0$,
the instantaneous, fractional mass loss  rate of a dark matter subhalo
is given  by $\dot{m}/m =  f(\psi)$ with $f(x)$ an  arbitrary function
($0  \leq x  \leq 1$),  to be  determined below.   In what  follows we
assume, for simplicity, that $f(x)$  is well described by a power-law,
and write
\begin{equation}
\label{mydecay}
\dot{m} = - {m \over \tau} \, \psi^{\zeta}
\end{equation}
Here $\tau$ is a characteristic time scale (in Gyr), and $\zeta$ is an
additional free  parameter that specifies  the mass dependence  of the
subhalo mass loss rate.  The negative sign is to emphasize that $m$ is
expected to  decrease with time.  For  a subhalo embedded  in a static
parent halo ($\dot{M}=0$) this yields
\begin{equation}
\label{myevolv}
m(t) = \left\{ \begin{array}{ll}
m_i \, {\rm exp}(-t/\tau) & \mbox{if $\zeta = 0$} \\
m_i \left[ 1 + \zeta \, \psi_i^\zeta \, (t/\tau) \right]^{-1/\zeta} &
\mbox{otherwise} 
\end{array} \right.
\end{equation}
where we  have used the  boundary condition $\psi(t=0) \equiv  \psi_i =
m_i/M$. 
\begin{figure*}
\centerline{\psfig{figure=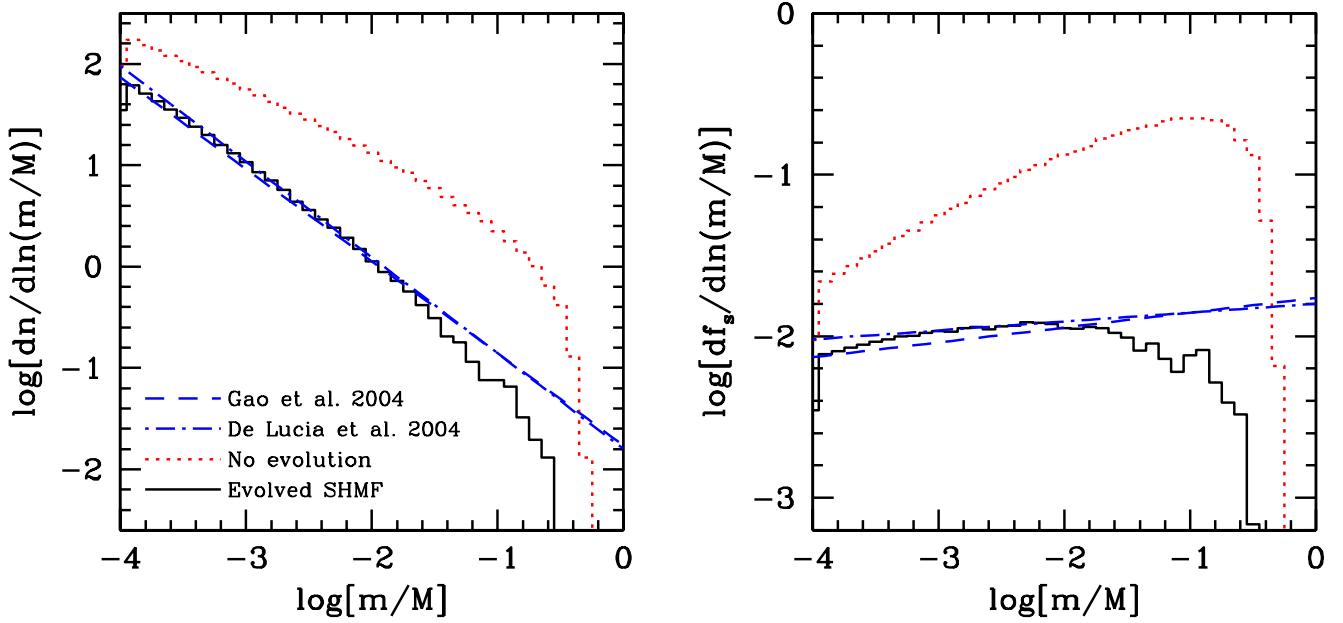,width=\hdsize}}
\caption{{\it Left-hand panel:} The solid histogram indicates the average, 
  evolved  SHMF for  a parent  halo with  $M_0=10^{15}  h^{-1} \Msun$,
  obtained  from  2000  merger   trees  with  $\tau_0=0.13  \Gyr$  and
  $\zeta=0.36$.  With these parameters the resulting SHMF best matches
  those of Gao \etal (2004) and De Lucia \etal (2004), shown as dashed
  and dot-dashed  lines, respectively. Note  that the model  reveals a
  high-mass  cut-off.  The  dotted histogram  indicates  the unevolved
  SHMF (i.e., without subhalo mass  loss) and is shown for comparison. 
  {\it Right-hand panel:} Same as left-hand panel, except that here we
  plot the  mass fraction  of dark matter  subhaloes. Note  that ${\rm
    d}f_s/{\rm d} {\rm ln}(m)$ of  the evolved subhaloes is very flat,
  indicating that low mass subhaloes contain a significant fraction of
  the total subhalo mass.}
\label{fig:comp}
\end{figure*}

We  emphasize at this  stage that  the power-law  form of  the average
mass-loss rate  has no  physical motivation. We  choose it  purely for
simplicity. Note  that $\dot{m}$  has to capture  the effects  of both
dynamical friction  and tidal  stripping.  One might  therefore expect
that the mass loss rates of individual subhaloes differs significantly
from the simple power-law form  adopted here.  However, recall that we
use $\dot{m}$ to  describe the {\it average} mass  loss rate, which is
not  necessarily of the  same form  as that  of individual  subhaloes. 
Furthermore, as we show below,  it does seem able to naturally explain
the  subhalo statistics  found in  numerical simulations,  without the
need for a more complicated functional form.  Nevertheless, a detailed
comparison against the average mass loss rates obtained from numerical
simulations is required  to check whether our power-law  form is truly
appropriate.

One  naturally expects  the  characteristic time  scale  $\tau$ to  be
related to the  dynamical time, $t_{\rm dyn}$, of  the parent halo. As
shown in Appendix~A,  in the idealized case of  homologous haloes, the
mass loss rate of a subhalo  on a circular orbit can indeed be written
in the  form~(\ref{mydecay}) with $\tau = t_{\rm  dyn}$. Since $t_{\rm
  dyn} \propto \rho^{-1/2}$,  and since the average density  of a dark
matter halo  is a function  of redshift, we  thus expect that  $\tau =
\tau(z)$.  The  average density  of a virialized  dark matter  halo at
redshift  $z$ is  given  by $\bar{\rho}(z)  =  \Delta_{\rm vir}(z)  \,
\rho_{\rm crit}(z)$. Here $\rho_{\rm crit}(z) = 3 H^2(z) / 8 \pi G$ is
the  critical density  for  closure, and  $\Delta_{\rm  vir}(z)$ is  a
cosmology-dependent quantity for which  we use the fitting function of
Bryan  \& Norman  (1998).   To  take proper  account  of the  expected
redshift dependence of the  characteristic time scale for subhalo mass
loss, we write
\begin{equation}
\label{mytau}
\tau = \tau(z) = \tau_0 \, 
\left( {\Delta_{\rm vir}(z) \over \Delta_{\rm vir}(0)}\right)^{-1/2} \, 
\left( {H(z) \over H_0} \right)^{-1}
\end{equation}
with $\tau_0$ a free  parameter that expresses the characteristic time
scale for subhalo mass loss at $z=0$.
\begin{figure*}
\centerline{\psfig{figure=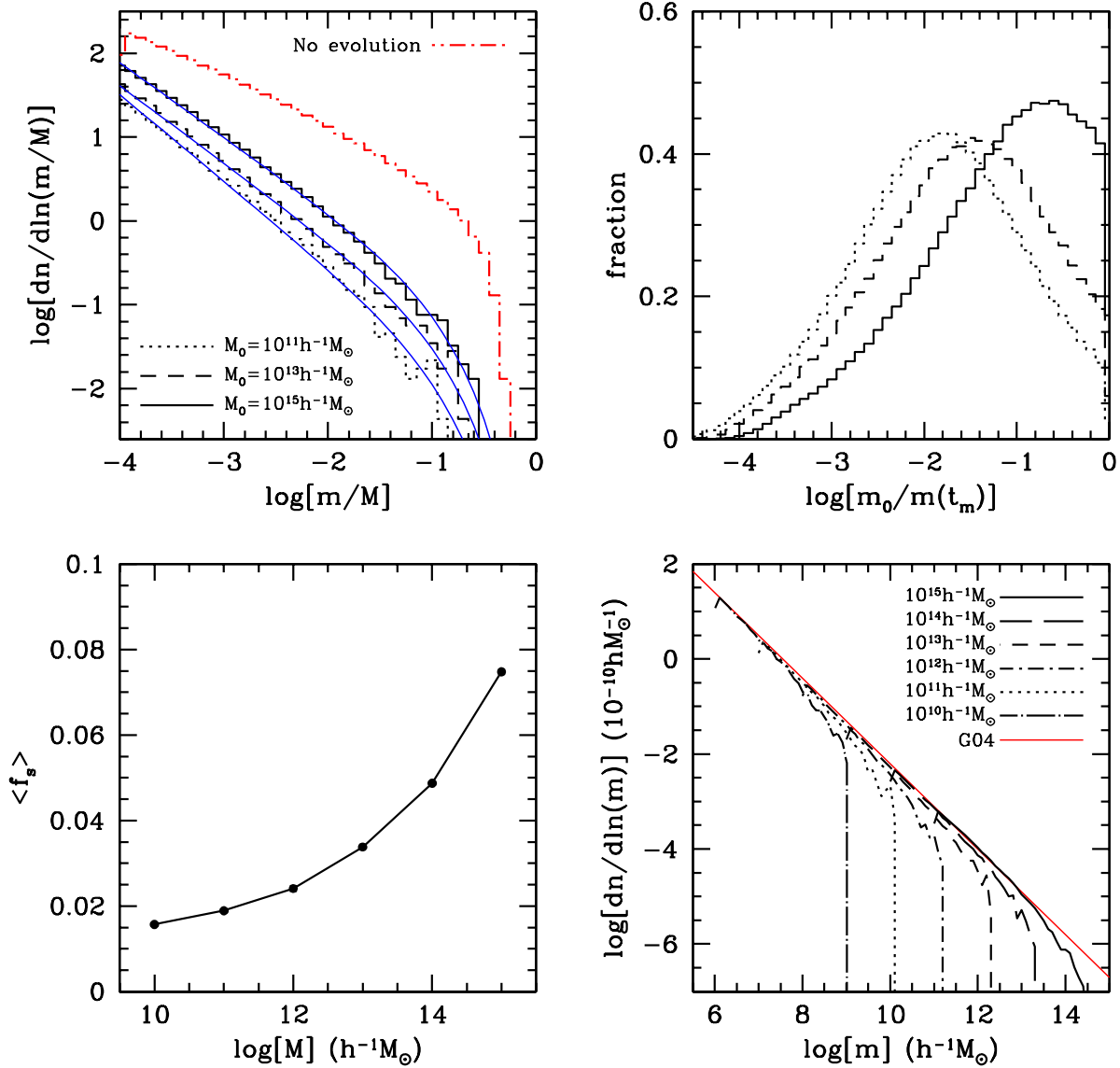,width=0.9\hdsize}}
\caption{Parent halo mass dependence of the SHMF.
  {\it Upper left-hand panel:} Average SHMF for parent haloes of three
  different masses,  as indicated. Less massive  parent haloes contain
  fewer subhaloes at any given  mass ratio $m/M$.  For comparison, the
  dot-dashed  histogram   indicates  the  unevolved   SHMF  (which  is
  virtually indistinguishable  for parent  haloes of different  mass). 
  The thin,  solid lines are  the SHMF fitting functions  described in
  Section~\ref{sec:fit}.  {\it  Upper right-hand panel:} Distributions
  of  the ratio  of present  day mass  to mass  at time  of accretion,
  $m_0/m(t_m)$, for all subhaloes in parent haloes with $M_0 = 10^{15}
  h^{-1}  \Msun$  (solid  histogram),  $M_0 =  10^{13}  h^{-1}  \Msun$
  (dashed  histogram),  and  $M_0  =  10^{11}  h^{-1}  \Msun$  (dotted
  histogram).  Note that subhaloes in more massive parent haloes have,
  on  average, lost  a smaller  fraction  of their  mass.  {\it  Lower
    left-hand panel:} The average  subhalo mass fraction, $\langle f_s
  \rangle$ (averaged  over 2000 merger  trees), as function  of parent
  halo mass: subhaloes in more  massive parent haloes contain a larger
  mass fraction.  {\it Lower  right-hand panel:} SHMFs scaled per unit
  parent halo mass.  Note that  with this scaling, the mass dependence
  is  completely  removed (except  for  the  high-mass cut-off,  which
  simply reflects  that $m/M<1$).  The  thin, solid line  labelled G04
  indicates  the results  (eq.~[\ref{gaofit}]) obtained  by  Gao \etal
  (2004) from  high-resolution,  numerical   simulations,  and  is  in
  excellent agreement with our model predictions.}
\label{fig:mass}
\end{figure*}

\subsection{Evolution of the population of subhaloes}
\label{sec:evolve}

Although  the subhalo  mass loss  rate in  a static  parent halo  is a
meaningful concept  from a physical  point of view, in  reality parent
haloes themselves  evolve due  to merging and  accretion. In  order to
take this  into account we utilize  the discrete time  stepping of our
merger trees.  At  the beginning of each time step  the parent halo is
assumed to increase its mass through (instantaneous) mergers ($\dot{M}
> 0$, $\dot{m}  = 0$), while during  the period in  between two merger
events  we  set   $\dot{M}  =  0$  and  evolve   $m(t)$  according  to
eq.~(\ref{myevolv}). The exact procedure is illustrated graphically in
Fig~\ref{fig:drawing}: At  $t=t_1$ halo  1 (with three  subhaloes) and
halo  2 (with  two subhaloes)  merge.  Since  $M_1 >  M_2$, halo  1 is
considered  the  new  parent halo,  with  halo  2  as a  subhalo.   In
addition,  the subhaloes of  $M_1$ are  preserved, and  are considered
subhaloes of the  new, merged halo.  The two  subhaloes of 2, however,
are  no longer considered  (i.e., we  do not  follow the  evolution of
sub-subhaloes).   From time  $t_1$ to  $t_2$, which  is when  the next
merging or  accretion event occurs, the subhaloes  evolve according to
our mass loss rate, i.e., eq.~(\ref{myevolv}) with $t=t_2-t_1$, $m_i =
m(t_1)$, and  $\tau = \tau(t_1)$.  This  procedure, hereafter referred
to as the `Monte-Carlo method',  yields, at each redshift, the evolved
SHMF.   In addition, we  also register  for each  subhalo the  time of
merging,  $t_m$, as  well as  its mass  at that  time,  $m(t_m)$.  The
abundance  of  these progenitor  haloes  as  function  of their  mass,
$m(t_m)$, is hereafter referred to as the unevolved SHMF.

In order to  calibrate our model, we tune  the free parameters $\zeta$
and $\tau_0$  such that  the SHMF of  parent haloes  with $M_0=10^{15}
h^{-1} \Msun$ matches  the subhalo mass functions of  Gao \etal (2004)
and De  Lucia \etal  (2004). Although these  two SHMFs,  obtained from
independent numerical simulations, are  very similar, the agreement is
not perfect. Including the results  of Tormen \etal (2004) we estimate
the accuracy of the absolute normalization to be about 20 percent, and
caution the reader  that the absolute normalization of  our results is
therefore  uncertain  by a  similar  amount.   Nevertheless, our  {\it
  relative} normalizations, which are the main topic of interest here,
should not be effected by  this.  A more robust absolute normalization
will have to await a larger sample of high-resolution simulations, and
a more detailed investigation of numerical resolution effects. 

The  dotted histogram  in the  left-hand panel  of Fig.~\ref{fig:comp}
plots the average {\it unevolved} SHMF obtained from 2000 merger trees
for a  parent halo of $M_0 =  10^{15} h^{-1} \Msun$, and  is shown for
comparison  with the evolved  SHMF (solid  histogram).  The  latter is
obtained from  the same 2000  merger trees using the  method described
above with  $\tau_0 = 0.13  \Gyr$ and $\zeta  = 0.36$.  These  are the
parameters  for which  we  obtain  the best-fit  to  the subhalo  mass
functions  of Gao \etal  (2004) and  De Lucia  \etal (2004),  shown as
dashed and dot-dashed curves,  respectively.  The agreement with these
SHMFs obtained from numerical simulations is very satisfactory, except
for a  high-mass cut-off in our  model, which is not  accounted for in
the simple power-law  fits to the published SHMFs  of Gao \etal (2004)
and  De  Lucia \etal  (2004).   Detailed  tests  have shown  that  the
location of  this high-mass cut-off  is robust to changes  in $\tau_0$
and/or   $\zeta$.    The  former   mainly   influences  the   absolute
normalization, while the latter controls the slope at small $m/M$.

The right-hand panel of Fig.~\ref{fig:comp} plots 
\begin{equation}
\label{subfrac}
{{\rm d}f_s \over {\rm d}\,{\rm ln}\,m} = 
\psi {{\rm d}n \over {\rm d}\,{\rm ln}\,m}
\end{equation}
which indicates  the {\it  mass} fraction of  the parent halo  that is
associated with subhaloes  of mass $m$.  Most of  the present day halo
mass  originates  from progenitors  (i.e.,  unevolved subhaloes)  with
$m(t_m) \sim  0.1 M_0$. In fact,  if one ignores  all progenitors with
$m(t_m)/M_0 <  10^{-4}$ one misses  only a negligible fraction  of the
entire  mass. This, however,  is not  the case  for the  {\it evolved}
SHMF.   Here ${\rm  d}f_s /  {\rm d}{\rm  ln}\,m$ is  remarkably flat,
indicating  that even the  evolved subhaloes  with $m/M  \leq 10^{-4}$
contribute  a  significant  fraction   of  the  total  subhalo  mass.  
Therefore,  it is  important to  always  indicate the  range of  $m/M$
considered when  quoting subhalo mass fractions, and  care is required
when comparing subhalo mass  fractions from different simulations with
different resolutions.

\section{The Mass Dependence of the Subhalo Mass Function}
\label{sec:massdep}

Having  calibrated our  mass-loss rate  by matching  the  subhalo mass
function obtained  from numerical simulations  for haloes with  $M_0 =
10^{15} h^{-1} \Msun$, we  now investigate what the Monte-Carlo method
predicts for different parent  halo masses. The upper, left-hand panel
of Fig.~\ref{fig:mass} plots the  SHMFs obtained using the Monte-Carlo
method  with  $\tau_0  = 0.13  \Gyr$  and  $\zeta  = 0.36$  for  three
different parent halo masses, as indicated (each SHMF is averaged over
2000  merger  trees).  For  comparison,  we  also  show the  unevolved
subhalo mass function, which is virtually identical for all three halo
masses. The evolved SHMFs,  however, are clearly mass-dependent with a
normalization that decreases systematically with decreasing halo mass.
These findings are  in good agreement with those  of Gao \etal (2004),
based on numerical  simulation that span three orders  of magnitude in
parent  halo mass,  and strongly  argue against  earlier claims  for a
universal  subhalo mass  function (e.g.,  Moore \etal  1998;  De Lucia
\etal 2004).

The mass dependence of the normalization of the evolved SHMF is simply
a reflection of  the fact that less massive  haloes form earlier, thus
providing more time for mass  loss to operate.  This is illustrated in
the  upper, right-hand  panel of  Fig~\ref{fig:mass}, which  plots the
distributions of $m_0/m(t_m)$ for parent haloes of $M_0=10^{11} h^{-1}
\Msun$   (dotted  histogram),   $M_0=10^{13}  h^{-1}   \Msun$  (dashed
histogram),  and $M_0=10^{15} h^{-1}  \Msun$ (solid  histogram). These
distributions,  clearly show  that  subhaloes in  less massive  parent
haloes have,  on average, lost  a relatively larger fraction  of their
mass  since they  were accreted.   The distributions  of $m_0/m(t_m)$,
however,  are  very broad;  while  some  subhaloes  have only  lost  a
negligible fraction  of their initial  mass (either because  they were
accreted relatively  late, or because they had  small, relative masses
to begin  with), others  have lost more  than $99.9$ percent  of their
mass since their time of accretion.

\subsection{The mass fraction in subhaloes}
\label{sec:massfrac}

To  quantify the  mass dependence  of the  SHMF we  consider  the mass
fraction of dark matter subhaloes with $m \geq 10^{-4} M$:
\begin{equation}
\label{sfrac}
f_s \equiv \int_{10^{-4}}^{1} \psi \, 
{{\rm d}n \over {\rm d}\psi} \, {\rm d}\psi
\end{equation}
The lower limit of $10^{-4}$ reflects the effective mass resolution of
our merger trees (see Section~\ref{sec:theory}). Throughout this paper
all subhalo  mass fractions therefore  only take account  of subhaloes
with masses above this resolution  limit. As discussed above, the {\it
  total} subhalo mass fraction can  easily be a factor two larger than
this.

The lower,  left-hand panel of Fig.~\ref{fig:mass}  plots $\langle f_s
\rangle$ at $z=0$ (where the  average is taken over 2000 merger trees)
as  function of  parent halo  mass. This  nicely illustrates  the mass
dependence of  the SHMF indicated above, namely  a systematic increase
of $\langle f_s  \rangle$ with increasing parent halo  mass.  From the
scale of  galaxy sized haloes  ($M_0 \simeq 10^{12} h^{-1}  \Msun$) to
that of massive  clusters ($M_0 \simeq 10^{15} h^{-1}  \Msun$) we find
that  $\langle f_s  \rangle$ increases  by  about a  factor three,  in
reasonable  agreement with  Gao \etal  (2004) and  Kang \etal  (2004). 
Note that $\langle  f_s \rangle(M)$ seems to asymptote  to a non-zero
value of  $\sim 0.01$ for  low mass parent  haloes, which is  simply a
reflection of the finite age of  the Universe; only in the limit where
the formation  time of  a halo  is infinitely long  ago will  there be
sufficient time to wipe out all substructure.
\begin{figure}
\centerline{\psfig{figure=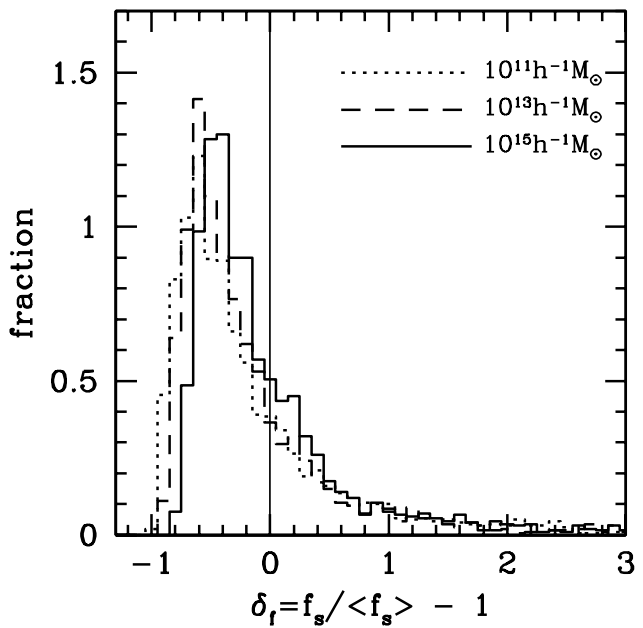,width=\hssize}}
\caption{Distributions of $\delta_f$ (eq.~[\ref{deltafs}]), obtained
  from 2000 independent merger  trees, for three different parent halo
  masses as  indicated. Note the  strong skewness, and  the relatively
  large dispersion. The vertical  line indicates the average for which
  $\delta_f=0$. See text for detailed discussion.}
\label{fig:scatter}
\end{figure}

\subsection{Subhalo mass functions per unit halo mass}
\label{sec:renorm}

The lower,  right-hand panel of Fig.~\ref{fig:mass}  shows the subhalo
mass  functions for six  different parent  halo masses  (each obtained
using $\tau_0=0.13  \Gyr$ and $\zeta  = 0.36$, and averaged  over 2000
merger trees), but this time normalized in a different way.  Following
Gao \etal (2004)  we divide the total number of  subhaloes in each bin
by  the total  mass of  all the  parent haloes  (in units  of $10^{10}
h^{-1} \Msun$)  to obtain the  subhalo abundance {\it per  unit parent
  halo mass}.  These abundances are  plotted as function of the actual
subhalo  mass, $m$,  rather than  the scaled  mass, $m/M$.   With this
particular  normalization,  the subhalo  mass  functions of  different
parent  halo masses  agree extremely  well (except  for  the high-mass
cut-off). The thin, straight line corresponds to
\begin{equation}
\label{gaofit}
{{\rm d}n \over {\rm d}m} = 10^{-3.2} \, 
\left( {m \over h^{-1}\Msun}\right)^{-1.9} h \Msun^{-1}
\end{equation}
which is the  best-fit subhalo abundance per unit  halo mass (ignoring
the high mass cut-off) obtained by  Gao \etal (2004).  As can be seen,
our results are in excellent agreement with those of Gao \etal (2004),
lending strong support for our simple model.
\begin{figure*}
\centerline{\psfig{figure=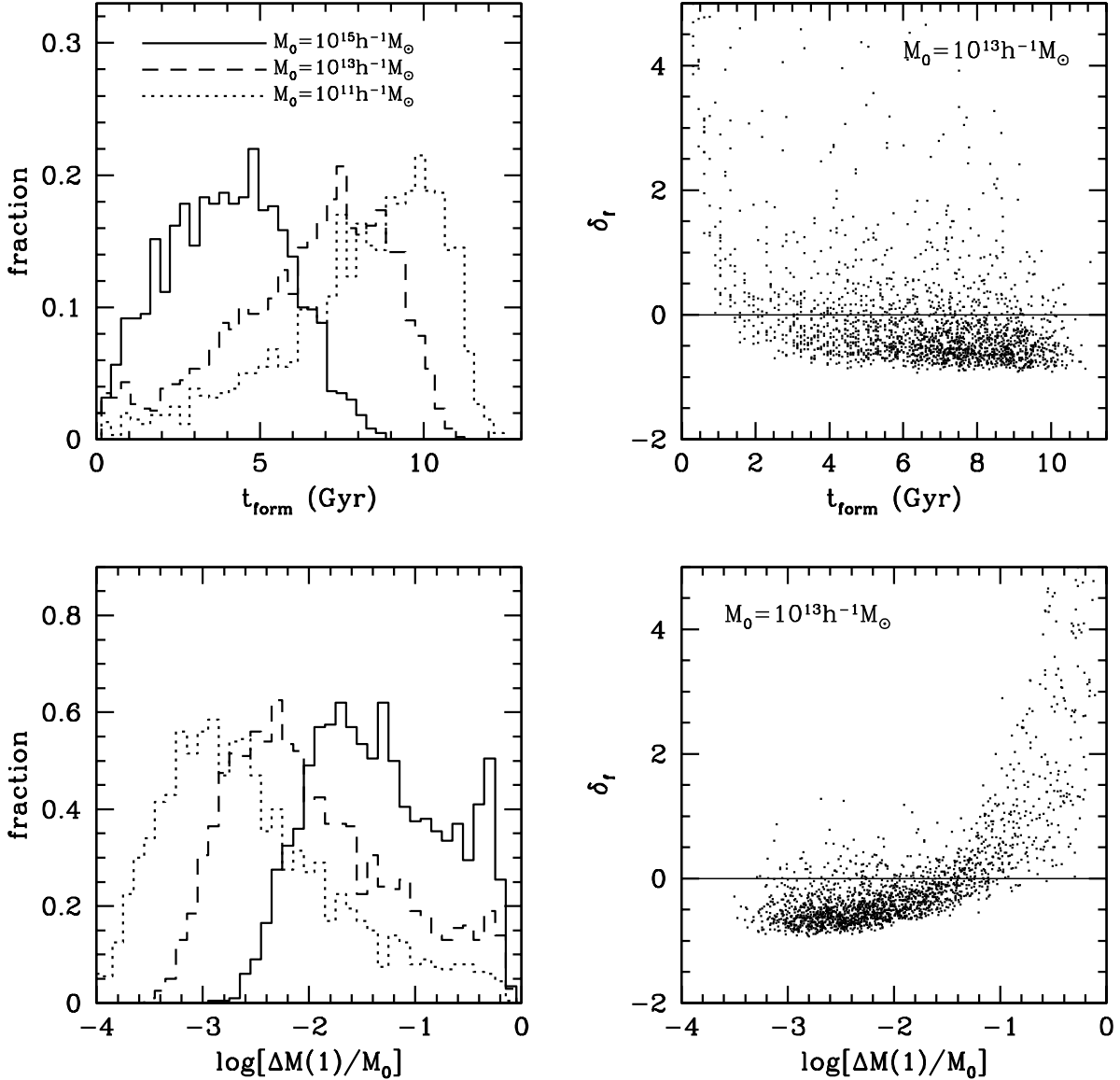,width=0.9\hdsize}}
\caption{Distributions of parent halo formation times, $t_{\rm form}$,
  (upper left-hand  panel) and of the mass  fraction $\Delta M(1)/M_0$
  accreted in  the last  1 Gyr (lower  left-hand panel).   Results are
  shown  for three  different parent  halo masses,  as  indicated. The
  right-hand  panels show,  for  a parent  halo  mass of  $M_0=10^{13}
  h^{-1} \Msun$,  how $t_{\rm  form}$ and $\Delta  M(1)/M_0$ correlate
  with $\delta_f$.   The relatively tight  relation between $\delta_f$
  and $\Delta  M(1)/M_0$ indicates that  the subhalo mass  fraction of
  individual haloes  depends mainly on their accretion  history in the
  last $\sim 1$ Gyr.}
\label{fig:his}
\end{figure*}
\begin{figure*}
\centerline{\psfig{figure=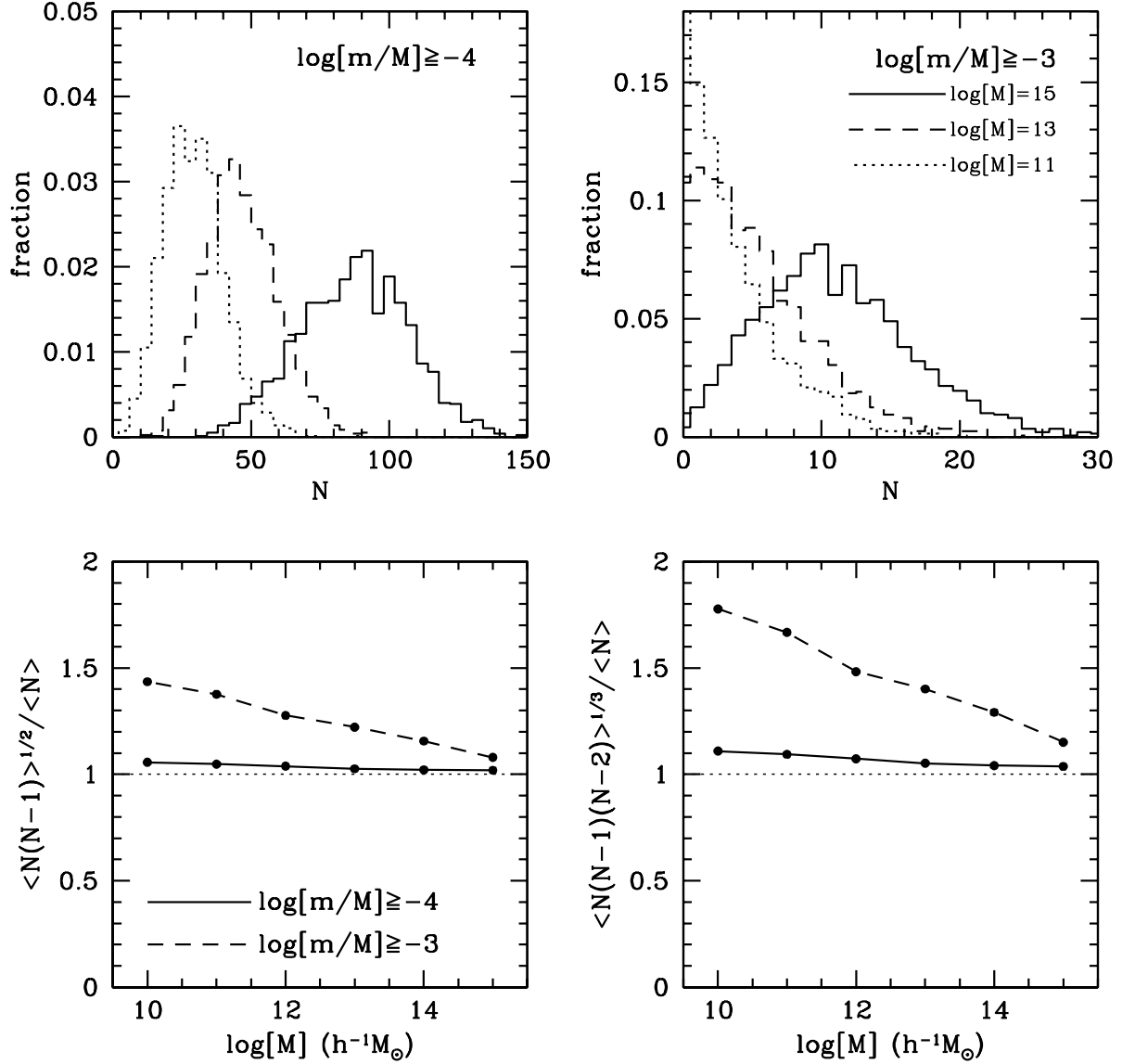,width=0.9\hdsize}}
\caption{The upper panels plot the distributions $P(N)$, where $N$ is
  the  number of  subhaloes with  $\log[\psi_0] \geq  -4$  (upper left
  panel) and $\log[\psi_0] \geq -3$ (upper right panel), respectively.
  Results,  obtained  from 2000  merger  trees,  are  shown for  three
  different parent  halo masses, as  indicated. The lower  panels plot
  $\langle N (N-1) \rangle^{1/2}/\langle N \rangle$ (lower left panel)
  and $\langle  N (N-1) (N-2) \rangle^{1/3}/\langle  N \rangle$ (lower
  right panel), which  express the second and third  moments of $P(N)$
  in  the  units of  that  of a  Poisson  distribution  with the  same
  $\langle N \rangle$. See text for a detailed discussion.}
\label{fig:poisson}
\end{figure*}

\section{Scatter in Subhalo Mass Functions}
\label{sec:scatter}

Thus far we  only focussed on the average SHMFs,  where the average is
taken  over all orbital  configurations and  over many  mass accretion
histories  (hereafter  MAHs).  However,  since  there is  considerable
scatter  in  MAHs  of  parent  haloes  of the  same  mass,  and  since
individual   haloes   may   have   significantly   different   orbital
distributions  for their  subhaloes,  one expects  a relatively  large
halo-to-halo variation in the SHMF. Here we use the Monte-Carlo method
to obtain an estimate of  this scatter.  Since this method implicitly
averages  over all  orbital configurations,  we can  only  address the
halo-to-halo scatter  due to variance  in the MAHs.  Our  estimates of
the amount of scatter are therefore to be considered lower-limits.

Fig.~\ref{fig:scatter} plots the distributions of
\begin{equation}
\label{deltafs}
\delta_f \equiv {f_s - \langle f_s \rangle \over \langle f_s
  \rangle}
\end{equation}
obtained from  2000 independent MAHs (merger trees).   Note that these
distributions  are extremely  skewed (not  surprising, given  that $f_s
\geq 0$),  and fairly  broad.  This indicates  that the  SHMF obtained
from a small  number of haloes, as is typically  the case with current
simulations,  may  not be  an  accurate  representation  of the  true,
average mass function.   This explains, at least partially,  why it is
so difficult  to infer from  numerical simulations whether or  not the
SHMF  depends  on  parent  halo   mass;  only  when  averaged  over  a
sufficiently large  number of parent  haloes will such a  trend become
evident.

The   upper,   left-hand  panel   of   Fig.~\ref{fig:his}  plots   the
distributions  of the  parent  halo formation  times, $t_{\rm  form}$,
defined as  the lookback time  at the redshift  where the mass  of the
most massive  progenitor reaches half the present  day mass.  Although
these distributions  are very broad,  there is a  clear mass-dependent
trend  in  that more  massive  haloes  form  later.  As  discussed  in
Section~\ref{sec:massdep},  this explains  why  $\langle f_s  \rangle$
increases  with parent  halo mass;  in  systems that  form later,  the
subhaloes have less time to  experience mass loss.  It therefore seems
natural that  the scatter  in $t_{\rm form}$  is the direct  source of
scatter in $f_s$. However, as evident from the upper, right-hand panel
of  Fig.~\ref{fig:his},  the correlation  between  $t_{\rm form}$  and
$\delta_f$ is, in fact, surprisingly weak.

We  find a must  stronger correlation  between $\delta_f$  and $\Delta
M(1)/M_0$ (lower right-hand panel of Fig.~\ref{fig:his}), with $\Delta
M(1)$ the mass that has been accreted by the parent halo in the last 1
Gyr.  This is easy to understand.  Since the characteristic time scale
for  subhalo mass  loss  is  relatively short  ($\tau_0  = 0.13$  Gyr)
compared  to  the  typical  halo  formation time,  the  subhalo  mass
fraction  in individual  systems is  dominated  by the  mass that  was
accreted  relatively  recently;  we  find  that  the  scatter  between
$\delta_f$ and  $\Delta M(t)  / M_0$ is  minimized for $t  \simeq 1.0$
Gyr,  which is  the  value  adopted here.   A  similar conclusion  was
reached by  Gao \etal  (2004) who found  that subhaloes  are typically
recent additions to their parent haloes, substantially more recent, in
fact,  than typical  dark  matter particles  (see  also ZB03).   Thus,
whereas the {\it average} SHMF depends strongly on formation time, the
SHMF of  an individual halo  simply reflects its accretion  history in
the last  $\sim 1$  Gyr. Although this  may seem contradictory,  it is
easy  to  understand when  looking  at  the  distributions of  $\Delta
M(1)/M_0$,   which   are   shown    in   the   lower-left   panel   of
Fig~\ref{fig:his} for  parent haloes of three different  masses. As is
evident,  more massive haloes  have, {\it  on average},  accreted more
mass recently, which reflects  their relatively later formation times,
and which is responsible for the mass-dependence of the average SHMF.
\begin{figure*}
\centerline{\psfig{figure=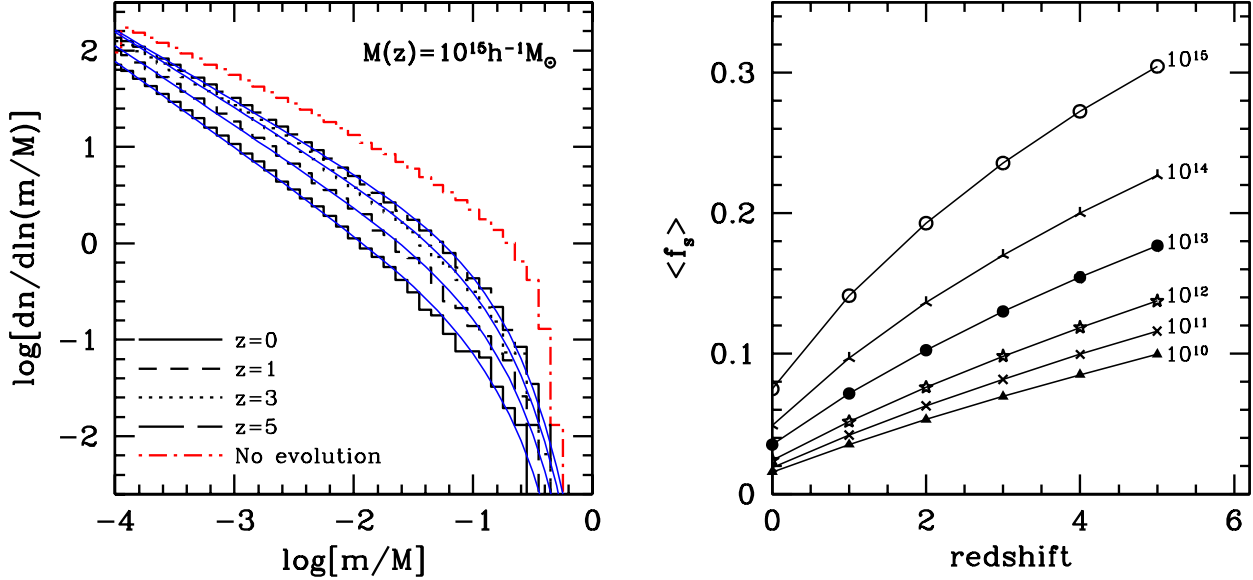,width=0.94\hdsize}}
\caption{Redshift dependence of the SHMF. {\it Left-hand panel:}
  Subhalo   mass  functions  for   parent  haloes   with  a   mass  of
  $M(z)=10^{15}  h^{-1}   \Msun$  at  four   different  redshifts,  as
  indicated. For  comparison, the  unevolved SHMF, which  is identical
  for all  four cases, is also  shown as a  dot-dashed histogram. Note
  that  haloes at high  redshift contain,  on average,  more subhaloes
  than  haloes  of  of  the  same  mass at  lower  redshifts.   As  in
  Fig.~\ref{fig:mass}  the  thin, solid  lines  are  the SHMF  fitting
  functions   described  in  Section~\ref{sec:fit}.   {\it  Right-hand
    panel:} Redshift dependence of  the average subhalo mass fraction,
  $\langle f_s \rangle$, for parent haloes of 5 different masses. Each
  curve is labelled by $M(z)$ (in $h^{-1} \Msun$).}
\label{fig:redshift}
\end{figure*}

In addition to the scatter in the subhalo mass fraction $f_s$, we also
investigate the scatter  in the {\it number} of  subhaloes.  The upper
two  panels of Fig.~\ref{fig:poisson}  plot the  distributions $P(N)$,
where  $N$ is  the number  of  subhaloes with  $\log[\psi_0] \geq  -4$
(upper  left panel) and  $\log[\psi_0] \geq  -3$ (upper  right panel),
respectively.   Results  are shown  for  three  different parent  halo
masses,  as  indicated.   As  above, these  probability  distributions
reflect  2000 independent  MAHs  (per parent  halo  mass).  These  two
panels, once again, clearly demonstrate that the subhalo mass function
is not  universal, but instead  depends strongly on parent  halo mass:
more   massive   haloes   contain   more  subhaloes   above   a   give
$\psi$-threshold.  The lower two panels plot ${\cal M}_2 \equiv\langle
N  (N-1)  \rangle^{1/2}/\langle  N  \rangle$, related  to  the  second
moment,   and    ${\cal   M}_3   \equiv   \langle    N   (N-1)   (N-2)
\rangle^{1/3}/\langle  N \rangle$,  related  to the  third moment,  of
these distributions, as  function of parent halo mass  $M$.  Note that
for a  Poissonian distribution  ${\cal M}_2 =  {\cal M}_3 =  1$, while
distributions   that   are   narrower  (sub-Poissonian)   or   broader
(super-Poissonian)  have  ${\cal   M}  <  1$  and  ${\cal   M}  >  1$,
respectively.    Clearly,   when   considering  all   subhaloes   with
$\log[\psi_0] \geq  -4$ (solid  lines), the $P(N)$  are very  close to
Poissonian.  However,  when only counting the  more massive subhaloes,
with $\log[\psi_0]  \geq -3$, the  distributions are super-Poissonian,
with a  clear trend  of increasing ${\cal  M}$ with  decreasing parent
halo mass. These results  are inconsistent with Kravtsov \etal (2004),
who finds that the number of subhaloes in numerical simulations follow
Poisson  statistics. This  may reflect  a generic  problem of  the EPS
formalism used here to construct the merger trees: as shown by various
authors  (Lacey \&  Cole 1993;  Somerville \etal  2000; van  den Bosch
2002; Wechsler \etal 2002), the  halo formation times predicted by EPS
are   systematically  offset  from   those  obtained   from  numerical
simulations. In particular, Somerville  \etal (2000) found the average
mass of  the largest  progenitor to be  larger with the  EPS formalism
than  in  the   simulations.   This  may  explain  why   we  find  the
non-Poissonian nature of $P(N)$ to be more pronounced for more massive
subhaloes.  Although   the  merger  trees   extracted  from  numerical
simulations  have their  own  problems, we  caution  that the  scatter
issues discussed here are probably less reliable than the average mass
trends.

\section{Redshift evolution of the Subhalo Mass Function}
\label{sec:redshift}

The  left-hand  panel  of  Fig.~\ref{fig:redshift} shows  the  average
subhalo mass  functions for  parent haloes of  the same mass,  $M(z) =
10^{15} h^{-1} \Msun$, but at different redshifts. At higher redshifts
parent haloes  of the same mass  have a larger  abundance of subhaloes
than their  counterparts at lower redshifts.  This  is quantified more
clearly  in  the right-hand  panel  of Fig.~\ref{fig:redshift},  which
shows the  redshift dependence of  the average subhalo mass  fraction. 
The various  curves are  labelled by the  parent halo mass  $M(z)$ (in
$h^{-1} \Msun$).  In all cases $\langle f_s \rangle(z)$ increases with
redshift, though  with a rate, ${\rm d}\langle  f_s \rangle/{\rm d}z$,
that  decreases  monotonically.  Roughly  speaking,  the subhalo  mass
fraction at $z=1$ is  about twice as large as that of  a halo with the
same mass at $z=0$.
\begin{figure}
\centerline{\psfig{figure=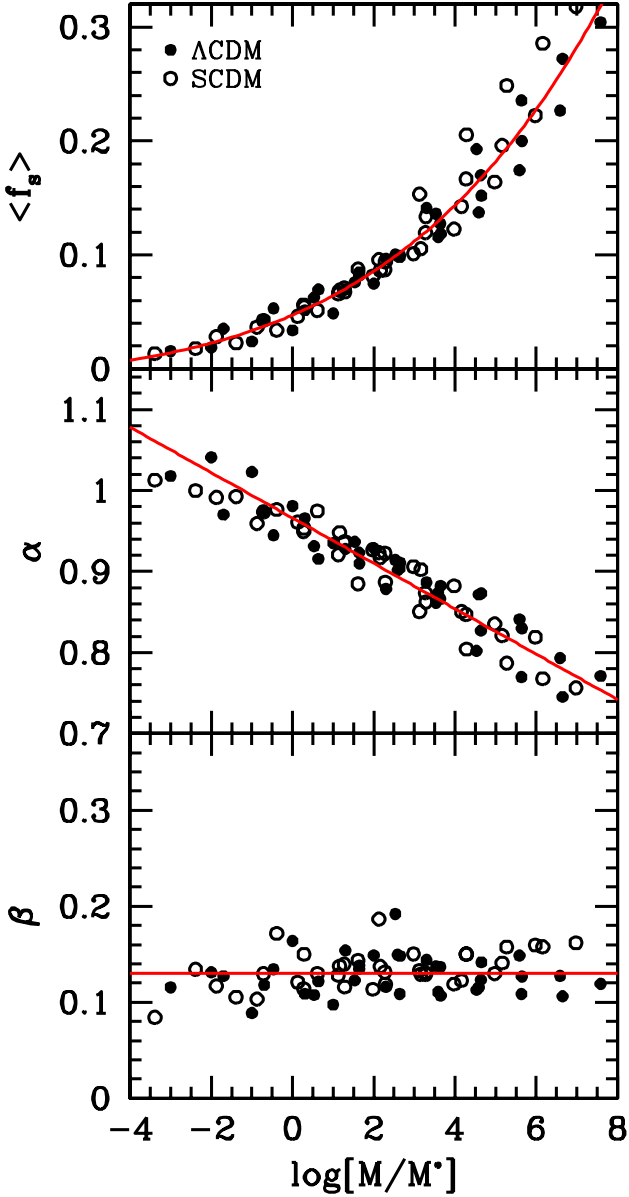,width=\hssize}}
\caption{The dependence of $\langle f_s \rangle$ (top panel), $\alpha$
  (middle  panel),  and  $\beta$  (bottom  panel) on  the  mass  ratio
  $M/M^{*}$.  Solid and open circles  correspond to parent haloes in a
  $\Lambda$CDM   concordance   cosmology   and   a   SCDM   cosmology,
  respectively.   Both  the  normalization  and the  shape  parameters
  $\alpha$  and   $\beta$  are  tightly   correlated  with  $M/M^{*}$,
  indicating  that the latter  is the  main parameter  determining the
  average SHMF of dark matter haloes. The solid lines are the best-fit
  functions~(\ref{fitfs}),~(\ref{fitalpha}),      and     $\beta=0.13$
  discussed in the text.}
\label{fig:res}
\end{figure}

The  subhalo mass  fraction  of a  given  halo at  redshift  $z$ is  a
trade-off  between  the  time  scale,  $t_{\rm  acc}$,  on  which  new
subhaloes are being `accreted' by the parent halo, and the time scale,
$\tau$, of  subhalo mass  loss.  The latter  evolves with  redshift as
described by eq.~(\ref{mytau}), and therefore was shorter in the past.
The former depends on the detailed MAH, and is thus a function of both
redshift and  parent halo mass.  In  the limit where  $t_{\rm acc} \ll
\tau$, subhalo  mass loss is  negligible and $f_s$ will  increase with
time.   The  opposite limit,  in  which  $t_{\rm  acc} \gg  \tau$,  is
equivalent to that  of subhalo mass loss in a  static parent halo.  In
this  case, $f_s$  will decrease  with  time. Since  the subhalo  mass
fraction always decreases  with time, the time scale  for subhalo mass
loss is  always smaller  than that of  mass accretion; $\tau  < t_{\rm
  acc}$. 

\section{An analytical fitting function for the subhalo mass function}
\label{sec:fit}

Since the unevolved SHMF is virtually independent of $M$, $z$, or even
of the  cosmological parameters (see Lacey  \& Cole 1993  and ZB03 for
detailed discussions), and since  we have adopted a universal, average
subhalo mass  loss rate, the  average, evolved SHMF simply  depends on
the halo formation  time.  This suggests that the  mass, redshift, and
cosmology dependence  of the average SHMF  can be written  as a simple
one-parameter  dependency  on  the  mass  ratio  $M/M^{*}$,  with  the
characteristic  non-linear mass $M^*(z)$  defined by  $\sigma(M^*,z) =
\delta_c(z)$.   Here  $\sigma^2(M,z)$  is  the mass  variance  of  the
smoothed density field at redshift $z$ and
\begin{equation}
\label{dc}
\delta_c(z) = 0.15 \, (12 \pi)^{2/3} \, [\Omega_m(z)]^{0.0055}
\end{equation}
is the critical threshold  for spherical collapse (e.g., Navarro \etal
1997).

The  top panel of  Fig.~\ref{fig:res} plots  the average  subhalo mass
fraction,  $\langle  f_s  \rangle$,  as  function  of  $M/M^*$  (solid
circles).   Results are  shown for  six different  parent  halo masses
(${\rm log}[M/h^{-1}\Msun] = 10,11,...,15$) at six different redshifts
($z=0,1,..5$). Although this yields values of $M/M^*$ up to $10^8$, we
caution that systems  with $M/M^* \gta 10^4$ are  extremely rare.  The
open  circles  indicate the  results  for  the  same halo  masses  and
redshifts,  but obtained  for  a SCDM  cosmology with  $\Omega_m=1.0$,
$\Omega_{\Lambda}=0.0$,   $h=0.5$,  and  $\sigma_8=0.7$.    All  these
different haloes follow a tight relation between $\langle f_s \rangle$
and $M/M^*$, which is well fitted by
\begin{equation}
\label{fitfs}
{\rm log}[ \langle f_s \rangle ] = \sqrt{0.4 ({\rm log}[M/M^*]+5)} - 2.74
\end{equation}
indicated by  the solid  line. This indicates  that, as  expected, the
average  subhalo mass  function is  completely specified  by  the mass
ratio $M/M^*$.   To quantify this  further we fit the  average subhalo
mass functions with a Schechter function of the form
\begin{equation}
\label{vo}
{{\rm d}n \over {\rm d}{\rm ln} \psi} = 
{\gamma \over \beta \, \Gamma(1-\alpha)}
\left( {\psi \over \beta} \right)^{-\alpha} \, {\rm exp}\left( 
-{\psi \over \beta} \right) 
\end{equation}
(cf. Vale \& Ostriker 2004).  Here $\beta M$ is a characteristic mass,
such that for $m \gg \beta M$ the SHMF reveals an exponential decline,
and $\gamma$ is the {\it total} subhalo mass fraction, i.e.,
\begin{equation}
\label{voint}
\gamma = \int_{0}^{\infty} \psi \, {{\rm d}n \over {\rm d}\psi} \, {\rm d}\psi
\end{equation}
Some   of   these  fits   are   shown   as   thin,  solid   lines   in
Fig.~\ref{fig:mass}       (upper       left-hand      panel)       and
Fig.~\ref{fig:redshift}  (left-hand panel), and  in general  match the
model SHMFs  extremely well.  Note that  in practice we  don't fit the
actual  SHMF,   ${\rm  d}n/{\rm   d}{\rm  ln}\psi$,  but   rather  the
corresponding ${\rm d}f_s/{\rm  d}{\rm ln}\psi$, treating $\alpha$ and
$\beta$  as  free parameters.   The  normalization,  $\gamma$, is  not
treated as  a free parameter, but  is fixed by requiring  to match the
subhalo mass fraction $f_s$, which implies
\begin{equation}
\label{norm}
\gamma = {f_s \over P(1-\alpha,1/\beta) - P(1-\alpha,10^{-4}/\beta)}
\end{equation}
with $P(a,x)$ the incomplete  Gamma function. 

The middle  and bottom panels of Fig.~\ref{fig:res}  plot the best-fit
$\alpha$  and $\beta$  for  each of  the  72 SHMFs  (2 cosmologies,  6
masses, 6  redshifts) as function of  the mass ratio  $M/M^*$. As with
the  average subhalo  mass  fraction, $\alpha$  and  $\beta$ are  both
tightly  correlated  with  the  parent  halo  mass  in  units  of  the
characteristic  non-linear mass; the  power-law slope  $\alpha$ scales
roughly linearly with ${\rm log}(M/M^*)$, which is best fit by
\begin{equation}
\label{fitalpha}
\alpha = 0.966 - 0.028 \, {\rm log}(M/M^*) \, ,
\end{equation}
while the parameter $\beta$ is best fit by $\beta = 0.13$.

We thus have  obtained an extremely simple recipe  to compute the {\it
  average} subhalo mass function for a parent halo of any mass, at any
redshift, and for any cosmology: compute the characteristic non-linear
mass  $M^{*}(z)$, and  use  eq.~(\ref{fitfs}) and~(\ref{fitalpha})  to
obtain  both  $f_s$ and  $\alpha$.   The  normalization $\gamma$  then
follows   from~(\ref{norm}),   which,   together  with   $\beta=0.13$,
completely specifies the SHMF.

\section{Conclusions}
\label{sec:concl}

We combined merger trees of  dark matter haloes, constructed using the
EPS formalism,  with a  simple prescription of  the average  mass loss
rate of dark  matter subhaloes, to compute subhalo  mass functions. We
calibrated the subhalo mass loss rate by matching the SHMFs of massive
haloes   with   $M_0   =   10^{15}   h^{-1}   \Msun$   obtained   from
high-resolution, numerical simulations. Under the assumption that this
average mass-loss rate only depends  on redshift and on the mass ratio
of sub- and parent halo, $m/M$, this method allows us to make detailed
predictions for the  mass and redshift dependence of  the SHMF, and to
investigate the halo-to-halo variance.

Our main conclusions are:
\begin{itemize}
  
\item Contrary  to previous claims,  the subhalo mass function  is not
  universal. Instead,  both the slope and the  normalization depend on
  the  ratio of parent  halo mass,  $M$, to  characteristic non-linear
  mass, $M^*$.  This simply reflects a halo formation time dependence,
  in which parent  haloes that form earlier have  a lower subhalo mass
  fraction,  because there is  relatively more  time for  subhalo mass
  loss to operate.
  
\item When the subhalo mass function  is normalized by the mass of the
  parent halo,  the abundance of subhaloes is  universal, in excellent
  agreement  with the numerical  simulations of  Gao \etal  (2004) and
  Kravtsov \etal (2004).
 
\item The  subhalo mass  function of an  individual halo  depends most
  strongly on  its accretion history in  the last $\sim  1$ Gyr.  This
  indicates, as  previously shown by  ZB03 and Gao \etal  (2004), that
  the population of the more  massive dark matter subhaloes is, at any
  time, relatively young.
  
\item  The  dependence  of   SHMF  on  the  recent  accretion  history
  introduces  a  large  halo-to-halo  scatter  in the  SHMFs,  with  a
  distribution  of  subhalo mass  fractions  that  is strongly  skewed
  towards large values.
 
\item The  average subhalo  mass function of  dark matter haloes  is a
  one-parameter family,  depending only on  the mass ratio  $M/M^{*}$. 
  We have provided simply fitting  functions that allow one to compute
  the average SHMF for a parent halo of any mass, at any redshift, and
  for any cosmology.

\end{itemize}

While this paper was being refereed, a paper appeared by Zentner \etal
(2004) which uses a similar semi-analytical  model as in ZB03 and TB04
to  compute  subhalo  statistics.    Using  a  proper  integration  of
individual orbits,  and taking detailed account  of dynamical friction
and  tidal stripping,  these  authors reach  conclusions  that are  in
excellent  agreement with  the simplified  method presented  here.  In
particular,  they find  that  (i)  the subhalo  mass  function is  not
universal but scales with halo mass, (ii) the subhalo mass function is
most  sensitive to  the most  recent accretion  history of  the parent
halo, and (iii) the distribution of the number of subhaloes per parent
halo   is  super-Poissonian.    The  good   agreement  of   this  more
sophisticated model with that presented here, provides further support
for our `orbit averaged' approach.

These results have a  number of important, astrophysical implications. 
For example, the prediction that  the average subhalo mass fraction in
galaxy  sized haloes  is a  factor three  lower than  in cluster-sized
haloes  has important implications  for the  magnitude of  the claimed
substructure crisis (Moore \etal 1999; Klypin \etal 1999b: D'Onghia \&
Lake 2004),  for the flux-ratio statistics of  multiply lensed quasars
(e.g., Chiba 2002; Metcalf \& Madau 2001; Brada\v{c} \etal 2002; Dalal
\& Kochanek 2002),  and for the build-up of  the galactic halo (Helmi,
White \&  Springel 2003). Furthermore, the redshift  dependence of the
subhalo mass  fraction impacts on the survival  probability of fragile
structures  in  dark  matter  haloes,  such as  tidal  streams  and/or
galactic disks (T\'oth  \& Ostriker 1992; Taylor \&  Babul 1991).  The
subhalo mass  functions derived here  may also be used  in combination
with the so-called halo model (see Cooray \& Sheth 2002 and references
therein) to give a full statistical description of the distribution of
dark matter  haloes down  to the level  of subhaloes. This  will proof
especially  fruitful in  combination with  the  conditional luminosity
function formalism  developed by Yang  \etal (2003) and van  den Bosch
\etal (2003), allowing for  a detailed, statistical description of the
relation  between light and  mass (see  also Vale  \& Ostriker  2004). 
Finally, the average mass loss  rates derived here may proof useful in
semi-analytical models for galaxy  formation, where a proper treatment
of the  evolution of subhaloes is extremely  important (Springel \etal
2002; Benson \etal 2002; Kang \etal 2004).

Finally we  point out that, although the  Monte-Carlo method presented
here is nice and simple, it  is important to be aware of its potential
shortcomings. For example, the  accuracy of the absolute normalization
of  our model  is only  as good  as  that of  the SHMFs  used for  its
calibration, which we  estimate to be about $\sim  20$ percent. In the
numerical  simulations used for  this calibration,  the masses  of the
parent haloes are defined as the masses within a sphere of density 200
times  the critical  density  at redshift  zero.   Therefore, we  have
implicitly  assumed that  the masses  in  the EPS  formalism used  to
construct our merger  trees are defined in the same  way.  Since it is
still unclear  what the proper  interpretation of these EPS  masses is
(see  White 2002  for a  detailed discussion),  this  `definition' may
introduce an  additional uncertainty in the  absolute normalization of
our results. The fact that the method to construct merger trees is not
without its own shortcomings (e.g.,  SK99; TB04; Sheth \& Tormen 1999;
Benson, Kamionkowski \& Hassani 2004) may have additional implications
for the  accuracy of  our results.  Furthermore,  we have  ignored the
weak dependence of halo concentration  on halo mass, which may cause a
(weak)  dependence of the  average subhalo  mass loss  rate on  $M$ in
addition  to $\psi$  (see e.g.,  ZB03). Finally,  we have  ignored any
possible effect due to subhalo-subhalo mergers. Although the excellent
agreement between  our results and  those of Gao \etal  (2004) suggest
that   none  of   these   effects  have   a   strong  impact,   large,
high-resolution  numerical simulations  are required  to  further test
both  the validity  of our  approach as  well as  the accuracy  of our
results.


\section*{Acknowledgements}

We  are  grateful  to  Gabriela  De Lucia,  Francesco  Haardt,  Savvas
Koushiappas, Andrey Kravtsov, Gao  Liang, Ben Moore, James Taylor, and
Andrew Zentner for useful discussions.



\appendix

\section[]{The Mass loss rate of subhaloes on circular orbits}
\label{sec:AppA}

Consider a subhalo with density distribution $\rho_s(r)$ on a circular
orbit  in a parent  halo with  density distribution  $\rho_p(r)$.  For
simplicity,  we  assume  that  both $\rho_s(r)$  and  $\rho_p(r)$  are
singular, isothermal spheres.

In the absence of tidal heating,  the mass loss rate of the subhalo is
given by
\begin{equation}
\label{dmdtsub}
{{\rm d}m \over {\rm d}t} = {{\rm d}m \over {\rm d}r_{\rm tid}} \,
{{\rm d}r_{\rm tid} \over {\rm d}t} 
\end{equation}
with $r_{\rm tid}$ the instantaneous  tidal radius of the subhalo.  We
define  this tidal  radius  as the  radius  of the  subhalo where  its
density is  equal to  that of  the parent halo  at the  orbital radius
$r_{\rm orb}$  of the satellite: $\rho_s(r_{\rm  tid}) = \rho_p(r_{\rm
  orb})$.  Since  $\rho_s(r)$ and  $\rho_p(r)$ are scale-free  we have
that
\begin{equation}
\label{drequal}
{1 \over r_{\rm tid}} \, {{\rm d}r_{\rm tid} \over {\rm d}t} = 
{1 \over r_{\rm orb}} \, {{\rm d}r_{\rm orb} \over {\rm d}t}
\end{equation}
The evolution of $r_{\rm orb}$  is governed by dynamical friction, and
is given by
\begin{equation}
\label{drorbdt}
{{\rm d}r_{\rm orb} \over {\rm d}t} = -0.428 \, {G \, m \over V_c \, r_{\rm
    orb}} \, {\rm ln}\Lambda
\end{equation}
(Binney  \& Tremaine  1987) with  $V_c$ the  circular velocity  of the
parent  halo and ${\rm  ln}\Lambda$ the  Coulomb logarithm,  which, to
leading order, is  just a function of the mass  ratio $m/M$ (Binney \&
Tremaine 1987).

Using that ${\rm d}m / {\rm  d}r_{\rm tid} = 4 \pi \rho_s(r_{\rm tid})
\, r_{\rm  tid}^2$ and $V^2_c = G  M / r_{\rm vir}$,  with $r_{\rm vir}$
the virial radius of the parent halo, we obtain that
\begin{equation}
\label{dmdts}
{{\rm d}m \over {\rm d}t} \propto - \rho_s(r_{\rm tid}) \, V_c \, 
{r_{\rm tid}^3 \, r_{\rm vir} \over r^2_{\rm orb}} \, {m \over M} \,
{\rm ln}\Lambda
\end{equation}
Defining the  dynamical time as $t_{\rm dyn}  \propto r_{\rm vir}/V_c$
and  using that  $m \propto  \rho_s(r_{\rm tid})  r_{\rm  tid}^3$, the
subhalo mass loss rate can be written as
\begin{equation}
\label{dmdtcirc}
{{\rm d}m \over {\rm d}t} \propto - {m \over t_{\rm dyn}} \,
\left( {r_{\rm vir} \over r_{\rm orb}}\right)^2 \, {m \over
  M} \, {\rm ln}\Lambda
\end{equation}
Therefore, when averaging over all possible circular orbits, i.e., all
possible  ratios  $r_{\rm  vir}/r_{\rm  orb}$,  one  obtains  an  {\it
  average} mass loss rate for which
\begin{equation}
\label{aver}
{{\rm d}m \over {\rm d}t} \propto - {m \over t_{\rm dyn}} \, g(m/M)
\end{equation}
with  $g(x)$ an arbitrary  function. This  is the  basic form  for the
average mass loss rate adopted  in this paper. Note that since $t_{\rm
  dyn}  \propto \rho^{-1/2}$, and  since the  average density  of dark
matter  haloes change  with  redshift, eq.~(\ref{aver})  automatically
implies a redshift dependence (see Section~\ref{sec:massloss}).

\label{lastpage}

\end{document}

%% file: macros_vdbosch.tex
\newcommand{\etal}{{et al.~}}

\newcommand{\kmsmpc}{\>{\rm km}\,{\rm s}^{-1}\,{\rm Mpc}^{-1}}

\newcommand{\Msun}{\>{\rm M_{\odot}}}

\newcommand{\Gyr}{\>{\rm Gyr}}

\newcommand{\beq}{\begin{equation}}
\newcommand{\eeq}{\end{equation}}

\def\gtsima{$\; \buildrel > \over \sim \;$}
\def\ltsima{$\; \buildrel < \over \sim \;$}
\def\prosima{$\; \buildrel \propto \over \sim \;$}
\def\gsim{\lower.7ex\hbox{\gtsima}}
\def\lsim{\lower.7ex\hbox{\ltsima}}
\def\simgt{\lower.7ex\hbox{\gtsima}}
\def\simlt{\lower.7ex\hbox{\ltsima}}
\def\simpr{\lower.7ex\hbox{\prosima}}
\def\la{\lsim}
\def\ga{\gsim}
\def\lta{\la}
\def\gta{\ga}

\newcommand{\apj}{ApJ}
\newcommand{\apjs}{ApJS}

\newcommand{\mnras}{MNRAS}
\newcommand{\aap}{A\&A}

\newdimen\hssize
\newdimen\hdsize
